\documentclass[11pt]{article}
\usepackage{graphicx}
\usepackage{epsfig}
\usepackage{amsmath}
\usepackage{amsthm}
\usepackage{amssymb}
\usepackage{rotating}
\usepackage{tabularx}
\usepackage{diagbox}
\usepackage{url}

\usepackage[round]{natbib}
\usepackage{color}

\usepackage{units}

\newcommand{\n}[1] {\mbox{\boldmath{$#1$}}}

\newcommand{\be}{\begin{eqnarray}}
\newcommand{\ee}{\end{eqnarray}}
\newcommand{\beq}[1]{\begin{equation}\label{#1}}
\newcommand{\eeq}{\end{equation}}
\newcommand{\ba}{\begin{eqnarray*}}
\newcommand{\ea}{\end{eqnarray*}}
\DeclareMathOperator*{\argmax}{arg\,max}

\newcommand{\tpack}[1]{{\textsf{#1}}}

\usepackage{vmargin}
\setmargins{2.5cm}       % left margin
{1.5cm}                  % top
{16.5cm}                 % width
{22.42cm}                % height
{10pt}                   % heading height
{1cm}                    % heading and text distance
{0pt}                    % footnotes height
{2cm}                    % footnotes and text distance

\title{Methods and Tools for Bayesian Variable Selection and Model Averaging in Univariate Linear Regression}
\author{Anabel Forte,\\ {\it Department of Statistics and Operations research, University of Valencia} \\ Gonzalo Garc\'ia-Donato \\ {\it Department of Economics and Finance, University of Castilla-La Mancha} \\and Mark F.J. Steel\footnote{Corresponding author: Mark Steel, Department of Statistics, University of Warwick, Coventry, CV4 7AL, UK; email: M.F.Steel@stats.warwick.ac.uk}\\ {\it Department of Statistics, University of Warwick}}

\date{}
\begin{document}
\maketitle

%\graphicspath{{../plots//}}

\begin{abstract}
%This paper focuses on uncertainty regarding variable selection in regression models, a problem that arises in many fields of application. 
In this paper we briefly review the main methodological aspects concerned with the application of the Bayesian approach to model choice and model averaging in the context of 
variable selection in regression models. This includes prior elicitation, summaries of the posterior distribution and computational strategies. We then examine and compare various publicly available {\tt R}-packages for its practical implementation summarizing and explaining the differences between packages and giving recommendations for applied users. We find that all packages reviewed lead to very similar results, but there are potentially important differences in flexibility and efficiency of the packages.
%Computational Methods

%R packages Guide

%User recommendations
\end{abstract}

\section{Motivation}
A very general problem in statistics is where several statistical models are proposed as plausible descriptions for certain observations $\n y$ and the observed data are used to resolve the model uncertainty. This problem is normally known as {\em model selection} or {\em model choice} if the aim is to select a single ``best'' model, but if the model uncertainty is to be formally reflected in the inferential process, we typically use  {\em model averaging}, where inference on issues that are not model-specific (such as prediction or effects of covariates) is averaged over the set of models under consideration.

%but, depending on the context and even on the discipline, may be introduced with alternative names like {\em model averaging} or {\em mixture modelling}.  In what follows we will refer to these problems as model selection (MS for short).
%such uncertainty are to be propagated through the inferential process.
%The ultimate goal in MS can be very assorted, ranging from selecting a single model (a question sometimes posed as: what is the true model?) to using all candidate models to produce more realistic estimations or predictions.

A particular important model uncertainty problem in practice is {\em variable selection} where the proposed models share a common functional form (eg. a normal linear regression model) but differ in which explanatory variables, from a given set, are included to explain the response. The focus in this paper will be on variable selection in the context of normal linear models, a problem formally introduced in Section \ref{sec:methodology}.

Model uncertainty is a classic problem in statistics that has been scrutinized from many different perspectives. Hence, quite often, the main issues for practitioners are to decide which methodology to use and/or how to implement the methodology in practice. One appealing approach is based on the Bayesian paradigm and is considered by many the {\em formal} Bayesian answer to the problem. This approach is the one based on the posterior probabilities of the models under consideration and results in a coherent and complete analysis of the problem and provides answers to practical questions. For instance, a single model can be selected as that most supported by the data (the model with the highest posterior probability) or inferences can be performed using the posterior model probabilities as weights, normally denoted by Bayesian model averaging (BMA).%, which was popularized in economics in the context of cross-country growth regression in \cite{FLS_JAE} and has since become a frequently used tool for dealing with uncertainty in economic problems (see {\it e.g.}~the recent special issue on ``Model Uncertainty in Economics'' published by the {\it European Economic Review} in 2016). 
In this paper we describe how the formal Bayesian method can be implemented in {\tt R} \citep{R}, analyzing the different packages that are currently available in CRAN ({\tt cran.r-project.org}). Emphasis is placed on comparison but also on putting in perspective the details of the implementations.

As with any Bayesian method, the prior distribution for the unknown parameters needs to be specified. It is well known that this aspect is particularly critical in model uncertainty problem since results are potentially highly sensitive to the priors used \citep[see e.g.][]{BergerPericchi01,LeySteel09}. In this paper, we pay  special attention to the family of priors in the tradition started by Jeffreys, Zellner and Siow \citep{Jef61,ZellSiow80,Zellner86} and continued by many other authors with important contributions during the last ten years. These types of priors, which we label {\em conventional}, are introduced in Section~\ref{pigamma}.  \cite{Baetal11} have recently shown that conventional priors have a number of optimal properties that make them a very appealing choice for dealing with model uncertainty.

\section{Bayesian variable selection in Linear Models}\label{sec:methodology}

Consider a Gaussian response variable $\n y$, size $n$, assumed to be explained by an intercept and some subset of $p$  possible explanatory variables with values grouped in the $n\times p$ matrix $\n X=(x_1,\dots,x_p)$. Throughout the paper we suppose that $n>p$ and that $\n X$ is of full column rank. We define a binary vector $\n\gamma=(\gamma_1,\dots,\gamma_p)^t$ where $\gamma_i=1$ if $x_i$ is included in the model $M_\gamma$ and zero otherwise. This is the variable selection problem, a model uncertainty problem with the following $2^p$ competing models:
\begin{equation}\label{theproblem}
M_\gamma:\n y=\alpha\n 1_n+\n X_\gamma\n\beta_\gamma+\n\varepsilon,
\end{equation}
where $\n\varepsilon\sim N_n(\n 0,\sigma^2\n I_n)$ and the $n\times p_\gamma$ design matrices $\n X_\gamma$ are all possible submatrices of $\n X$.
If we choose the null matrix for $X_\gamma$, corresponding to $\gamma=\n 0$, we obtain the null model with only the intercept
\begin{equation}\label{null}
M_0 : \n y=\alpha\n 1_n+\n\varepsilon.
\end{equation}
Without loss of generality, we assume that columns of $\n X$ have been centered on their corresponding means, which makes the covariates orthogonal to the intercept, and gives the intercept an interpretation that is common to all models. The set of all competing models is called the model space and is denoted as ${\cal M}$.

Assuming that one of the models in ${\cal M}$ is the true model, the posterior probability of any model is
\begin{equation}\label{postprob}
Pr(M_{\gamma^*}\mid\n y)=\frac{m_{\gamma*}(\n y)Pr(M_{\gamma^*})}{\sum_\gamma m_\gamma(\n y)Pr(M_\gamma)},
\end{equation}
where $Pr(M_\gamma)$ is the prior probability of $M_\gamma$ and $m_\gamma$ is the integrated likelihood with respect to the prior $\pi_\gamma$:
\begin{equation}\label{marg}
m_\gamma(\n y)=\int\, p_\gamma(\n y\mid \n\beta_\gamma,\alpha,\sigma)\, \pi_\gamma(\n\beta_\gamma,\alpha,\sigma^2)\, d\n\beta_\gamma\,d\alpha\,d\sigma^2,
\end{equation}
also called the (prior) marginal likelihood.  Note that, for $\gamma=0$ this integrated likelihood becomes:
\begin{equation}\label{marg0}
m_0(\n y)=\int\, p_0(\n y\mid \alpha,\sigma)\, \pi_0(\alpha,\sigma^2)\,d\alpha\,d\sigma^2,
\end{equation}

An alternative expression for (\ref{postprob}) is based on the Bayes factors:
\begin{equation}\label{postprob2}
Pr(M_{\gamma^*}\mid\n y)=\frac{B_{\gamma^*}(\n y)Pr(M_{\gamma^*})}{\sum_\gamma B_\gamma(\n y)Pr(M_\gamma)},
\end{equation}
where $B_\gamma$ is the Bayes factor of $M_\gamma$ to a fixed model, say $M_0$ (without any loss of generality) and hence $B_\gamma=m_\gamma/m_0$ and $B_0=1$. %\rojo{**why do we need to separately state $M_0$ in (3) and (5)?***}\azul{done}

The prior on the model parameters implicitly assigns posterior point mass at zero for those regression coefficients that are not included in $M_{\gamma}$,
which automatically induces sparsity. %Other ways of inducing sparsity is through the so called spike and slab priors. Nevertheless, the version of these priors that is normally implemented is not a formal Bayesian method as the posterior distribution (\ref{postprob}) does not exist.

As stated in the introduction, we are mainly interested in software that implements the formal Bayesian answer which implies that we use the posterior distribution in (\ref{postprob}). Even with this important characteristic in common there could be substantial differences between {\tt R}-packages (leaving aside for the moment details on programming and the interface used) due to the following three aspects:
\begin{itemize}
  \item the priors that the package accommodates, that is, $\pi_\gamma(\n\beta_\gamma,\alpha,\sigma^2)$ and $Pr(M_\gamma)$,
  \item the tools provided to summarize the posterior distribution and obtain model averaged inference,
  \item the numerical methods implemented to compute the posterior distribution.
\end{itemize}

We now succinctly revise the main methodological proposals for the above points. Emphasis is placed in presenting the revision in a way that accommodates the methods implemented in the different {\tt R} packages.

\subsection{Prior Specification}\label{pigamma}
The two inputs that are needed to obtain the posterior distribution are $\pi_\gamma$ and $Pr(M_\gamma)$: the $2^p$ prior distributions for the parameters within each model and the prior distribution over the model space, respectively.

Without loss of generality, the prior distributions $\pi_\gamma$ can be expressed as
$$
\pi_\gamma(\n\beta_\gamma,\alpha,\sigma^2)=\pi_\gamma(\n\beta_\gamma\mid\alpha,\sigma^2)\pi_\gamma(\alpha,\sigma^2).
$$

Under the conventional approach \citep{FLS01} the standard Jeffreys' prior is used for the parameters that are common to all models
\begin{equation}\label{priorcommon}
  \pi_\gamma(\alpha,\sigma^2)=\sigma^{-2}
\end{equation}
and for $\pi_\gamma(\n\beta_\gamma\mid\alpha,\sigma^2)$ we adopt either a normal or mixtures of normal distributions centered on zero \citep[``by reasons of similarity''][]{Jef61} and scaled by $\sigma^2(\n X^t_\gamma\n X_\gamma)^{-1}$ \citep[``a matrix suggested by the form of the information matrix''][]{ZellSiow80} times a factor $g$, normally labelled as ``$g$-prior''. Recent research has shown that such conventional priors possess a number of optimal properties that can be extended by putting specific priors on the hyperparameter $g$. Among these properties are invariance under affine transformations of the covariates, several types of predictive matching and consistency \citep[for details see][]{Baetal11}.

The specification of $g$ has inspired many interesting studies in the literature. Of these, we have collected the most popular ones in Table~\ref{gTable}. %This high diversity is a first reason for the the existence of the different {\tt R}-packages.

\begin{table}[t!]
\begin{center}
{\small\scalebox{0.8}{
\begin{tabular}{lllc}
Proposal & Reference & Name  & Label\\
\hline
\multicolumn{4}{l}{{\bf Constant $g$}}\\
\hline
$g=n$ & \cite{Zellner86, KassWass95} & Unit Information prior (UIP) & C1\\
$g=p^2$ & \cite{FosGeo94} & Risk inflation criterion prior (RIC) & C2\\
$g=\max\{n,p^2\}$ & \cite{FLS01} & Benchmark prior (BRIC) & C3\\
$g=\log(n)$ & \cite{FLS01} & Hannan-Quinn (HQ) & C4\\
$g_\gamma=\hat{g}_\gamma$ & \cite{liang08} & Local Empirical Bayes (EBL) & C5\\
\hline
\multicolumn{4}{l}{{\bf Random $g$}}\\
\hline
$g\sim IGa(1/2,n/2)$ & \cite{Jef61,ZellSiow80,ZellSiow84} & Cauchy prior (JZS) & R1\\
$g|a\sim\pi(g)\propto (1+g)^{-a/2}$ & \cite{liang08} & hyper-g  & R2\\
$g|a\sim\pi(g)\propto (1+g/n)^{-a/2}$ & \cite{liang08} & hyper-g/n  & R3\\
$g\sim \pi(g)\propto (1+g)^{-3/2},\,g>\frac{1+n}{p_\gamma+1}-1$ & \cite{Baetal11} & Robust prior  & R4\\
\hline
\end{tabular}
}}
\end{center}\caption{\small Specific proposals for the hyperparameter $g$ in the literature. Column ``Label'' will be used as convenient reference to particular proposals throughout the paper. For the priors on $g$, $a>2$ to ensure a proper prior and $p_\gamma$ denotes the number of covariates in $M_\gamma$.}\label{gTable}
\end{table}

%\rojo{Notice that, in the definition of conventional priors,  $\n X^t_\gamma\n X_\gamma$  should be inverted for every model $M_\gamma$ and hence it should have full rank even for the model containing all the potential covariates. This is only possible if the number of data $n$ is larger than the number of considered covariates $p$ so for the rest of the paper $n>p$ will be assumed. }

Related with the conventional priors is the proposal by \cite{Raf95} which is inspired by asymptotically reproducing the popular Bayesian Information Criterion \citep{Sch1978}. \cite{Raf95} proposes using the same covariance matrix as the Unit Information Prior (see Table~\ref{gTable}) but with mean the maximum likelihood estimator $\hat{\n\beta_\gamma}$  (instead of the zero mean of the conventional prior).

Other priors specifically used in model uncertainty problems are the spike and slab priors, that assume that the components of $\n\beta$ are independent, each having a mixture of two distributions: one highly concentrated on zero (the spike) and the other one quite disperse (the slab). There are two different developments of this idea in the literature. In the original proposal by \cite{MitBea88} the spike is a degenerate distribution at zero so this fits with what we have called the formal approach. The proposal by \cite{GeoMc93} in which the spike is a continuous distribution with a small variance also received a lot of attention, perhaps for computational advantages. In this implementation there is no posterior distribution over the model space as every model smaller than the full model has zero probability.

With respect to the priors over the model space ${\cal M}$, a very popular starting point is
\begin{equation}\label{eq:MSprior}
Pr(M_{\gamma}\mid\theta)=\theta^{p_\gamma}(1-\theta)^{p-p_\gamma},
\end{equation}
where $p_\gamma$ is the number of covariates in $M_\gamma$, and the hyperparameter $\theta\in(0,1)$ has the interpretation of the common probability that a given variable is included (independently of all others).

Among the most popular default choices for $\theta$ are
\begin{itemize}
  \item Fixed $\theta=1/2$, which assigns equal prior probability to each model, i.e $Pr(M_\gamma)=1/2^p$;
  \item Random $\theta\sim\mbox{Unif}(0,1)$, giving equal probability to each possible number of covariates or model size.
\end{itemize}

Of course many other  choices for $\theta$ -- both fixed and random--  have been considered in the literature. In general, fixed values of $\theta$ have been shown to perform poorly in controlling for multiplicity (the occurrence of spurious explanatory variables as a consequence of performing a large number of tests) and can lead to rather informative priors. This issue can be avoided by using random distributions for $\theta$ as, for instance, the second proposal above that has been studied in \cite{ScottBerger09}. Additionally, \cite{LeySteel09} consider the use of $\theta\sim$Beta$(1,b)$ that results in a binomial-beta prior for the number of covariates in the model or the model size, $W$:
$$
Pr(W=w\mid b)\propto {p \choose w}\Gamma(1+w)\Gamma(b+p-w), \,\, w=0,1,\ldots,p.
$$
Notice that for $b=1$ this reduces to the uniform prior on $\theta$ and also on $W$. As \cite{LeySteel09} highlight, this setting is useful to incorporate prior information about the mean model size, say $w^\star$. This would translate into $b=(p-w^\star)/w^\star$.

\subsection{{Summaries of the posterior distribution and model averaged inference}}\label{Sec:Summaries}
%Whatever the method used the results need to be summarized and reported. As we are working in the context of a fully Bayesian analysis the main summary should be related to the study of posterior distributions.

%When solving a problem of variable selection, and as shown before,  we can obtain posterior distributions for model specific parameters such as regression coefficients and posterior probabilities for each model.

%Summarizing posterior distributions in the parametric space needs some caution given the that the meaning of a parameter could considerably vary among models. In this sense it make sense to look into the posterior distribution of a parameter calculated as a weighted average of its posterior distribution under each model but, given its (plausible) multi-modality, measures of central tendency do not seem sensible here.

The simplest summary of the posterior model distribution (\ref{postprob}) is its mode
$$
\argmax_\gamma\, Pr(M_\gamma\mid\n y).
$$
This model is the model most supported by the information (data and prior) and is normally called the HPM (stands for highest posterior model) or MAP (maximum a posteriori) model. Clearly, a measure of uncertainty regarding this summary is reflected by its posterior probability which should always be reported.

When $p$ is moderate to large, posterior probabilities of individual models can be very small and their interpretation loses appeal. In such situations, posterior inclusion probabilities (normally denoted as PIP) are very useful.
\begin{equation}
\label{eq:inclprob}
Pr(\gamma_i=1 \mid \n y)= \sum_{x_i \in M_\gamma} Pr(M_\gamma\mid \n y).
\end{equation}
These should be understood as the importance of each variable for explaining the response. Interestingly, these probabilities are used to define another summary, namely the median probability model (MPM) which is the model containing the covariates with inclusion probability larger than $0.5$. This model is studied in \citep{BarBer04} and they show that, in some situations, it is optimal for prediction.

Extending the idea of inclusion probabilities, it is interesting to obtain measures of joint importance of sets of regressors on the response. For instance, we can compute the posterior probability of two (or more) covariates  occurring together in the model or the probability that a covariate enters the model given that another covariate is already present (or not). These quantities are known as joint posterior probabilities and conditional posterior probabilities, respectively, and are studied, with other related summaries, in \cite{LeySteel07} (and references therein).

A measure of the model complexity is given by
\begin{equation}
\label{eq:ppdim}
Pr(W=w\mid \n y)= \sum_{M_\gamma: p_\gamma=w} Pr(M_\gamma\mid \n y),
\end{equation}
which is the posterior probability mass function of the model size.

{The posterior distribution easily allows for obtaining model averaged estimates of any quantity of interest $\Lambda$ (assuming it has the same meaning across all models). Suppose $\hat{\Lambda}_\gamma$ is the estimate of $\Lambda$ you would use if $M_\gamma$ were the true model. Then the model averaged estimate of $\Lambda$ is
\begin{equation}
\label{eq:bma}
\hat{\Lambda}=\sum_{M_\gamma}\, \hat{\Lambda}_\gamma\, Pr(M_\gamma\mid \n y),
\end{equation}
which has the appeal of incorporating model uncertainty. }

{When $\Lambda$ refers to regression coefficients ($\beta_i$) the model averaged estimates should be used and interpreted with caution as they could be potentially misleading since the `same' parameter may have a different meaning in different models \citep{BergerPericchi01}. Also the posterior distribution of $\beta_i$ is a discrete mixture and hence summaries like the mean are not natural descriptions.}

{One particular appealing application of this technique is in predicting new values $y^\star$ of the dependent variable  associated with certain values of the covariates. In this case $\Lambda$ could be the moments of $y^\star$ or even the whole predictive distribution. Apart from their intrinsic interest, predictions can be a very useful tool to run predictive checks (often using score functions) {\it e.g.}~to compare various prior specifications.
}

\subsection{Numerical methods}\label{secNM}
There are two main computational challenges in solving a model uncertainty problem. First is the integral in (\ref{marg}) and second is the sum in the denominator of (\ref{postprob}) which involves many terms if $p$ is moderate or large.

Fortunately, in normal models, conventional priors combine easily with the likelihood, and conditionally on $g$ lead to closed forms for $m_\gamma(\n y)$. Hence, at most, a univariate integral needs to be computed when $g$ is taken to be random. Interesting exceptions are the Robust prior of \cite{Baetal11} and the prior of \cite{MarGeo10}, which despite assuming a hyper prior on $g$ induce closed form marginals. This is done by making the prior on $g$ dependent on the size of the model considered, either through the hyperparameters or through truncation.

The second problem, related with the magnitude of the number of models in {$\cal M$} (i.e.~$2^p$), could be a much more difficult one. If $p$ is small (say, $p$ in the twenties at most) exhaustive enumeration is possible but if $p$ is larger, heuristic methods need to be implemented. The question of which method should be used has been studied in \cite{Ga-DoMa-Be13} which classify strategies as i) MCMC methods to sample from the posterior (\ref{postprob}) in combination with estimates based on frequencies and ii) searching methods looking for `good' models with estimates based on renormalization (i.e with weights defined by the analytic expression of posterior probabilities, cf.~(\ref{postprob}). They show that i) is potentially more precise than ii) which could be biased by the searching procedure. Approach i) is the most standard approach but different implementations of ii) have lead to fruitful contributions.
%\rojo{**I made some changes here: the SVSS is often used the approach with the prior of \cite{GeorgeMcCulloch93}, which is necessarily in class (i), so I think it might be confusing**}
%Among these, we highlight the Stochastic Search Variable Selection (SSVS) of \cite{GeorgeMcCulloch97}; the
The proposals in \cite{Rafteryetal97} and \cite{FLS01} which are based on a Metropolis-Hasting algorithm called $MC^3$ (originally introduced in \cite{MadYork95}) could be in either class above, while the implementation in \cite{eicheretal11} based on a leaps and bound algorithm proposed by \cite{Raf95} is necessarily in (ii), since model visit frequencies are not an approximation to model probabilities in this case.

\section{\texttt{CRAN} packages screening}
In what follows we will write the name of the packages using the font \tpack{package}; functions as {\tt function()} and arguments as {\tt argument}.

We seek in CRAN all possible packages that, potentially, could be used to implement the Bayesian approach to variable selection. The key words used to search in \texttt{CRAN} were {\em Model Selection}, {\em Variable Selection}, {\em Bayes Factor} and {\em Averaging}. The last search was on June 26, 2015 and we found a total of 13 packages: \tpack{ VarSelectIP}; \tpack{ spikeslab} \citep{Ishwaetal13}; \tpack{ spikeslabGAM} \citep{MorRou15}; \tpack{ ensembleBMA} \citep{Fraetal15}; \tpack{ dma} \citep{McCoretal14}; \tpack{ BMA} \citep{RafHoetal15}; \tpack{ mglmn} \citep{KaNa15}; \tpack{varbvs} \citep{CarSte12}; \tpack{INLABMA} \citep{BiGoRue15}; \tpack{BayesFactor} \citep{MorRou15}; \tpack{BayesVarSel} \citep{GarFor15}; \tpack{BMS} \citep{FeldZeu15} and \tpack{mombf}\citep{Rossell14}.

%\rojo{**should we also consider R2GUESS (available at \\ http://CRAN.R-project.org/package=R2GUESS and described in a January 2016 JSS paper by Liquet et al. doi: 10.18637/jss.v069.i02)? It is for multivariate y and we only focus on their $q=1$, but it still seems a relevant contender, especially as it uses more complex numerical methods and is essentially wrapped C++ code?** }\azul{Better after having an almost-final version of the paper or even during the reviewing process as this is something referees will ask for.}

From these, \tpack{VarSelectIP}, appeared as not longer supported and, within the rest, only the last four implement conventional priors described in the previous section to perform variable selection in linear models and hence will be considered for detailed description and comparison in the following sections. Particularly, \tpack{BayesVarSel} and \tpack{BMS} seem to be specifically conceived for that task, while the main motivation in \tpack{BayesFactor} and \tpack{mombf} seem different. \tpack{BayesFactor} provides many interesting functionalities to carry out $t$-tests, ANOVA-type studies and contingency tables using (conventional) Bayes factors with special emphasis on the specification of the hyper parameter $g$ for certain design matrices. On the other hand, \tpack{mombf} focuses on a particular type of priors for the model parameters, namely the non-local priors \citep{JohRos10,JohRos12}, applied to either the normal scenario considered here or probit models.

Of the other packages we found, \tpack{spikeslab} and \tpack{spikeslabGAM}, implement spike and slab priors in the spirit of the approach by \cite{GeoMc93} and hence are not directly comparable with packages that compute the posterior distribution over the model space. Interestingly, the original spike and slab approach by \cite{MitBea88} is used as the base methodology in \tpack{varbvs} but with a specific development by
\cite{CarSte12} with extreme high dimensional problems ($p>>$) in mind. Finally, \tpack{BMA} provides the posterior distribution over the model space, but based on the BIC criterion.

Some other packages consider statistical models that are not of the type studied here (linear regression models). This is the case for \tpack{ensembleBMA}, which implements BMA for weather forecasting models and \tpack{dma} which focuses on dynamic models.

\tpack{INLABMA} interacts with the INLA \citep{RueMarCho09} methodology for performing model selection within a given list of models. The priors there used are those in the {\tt R} package INLA which are not model selection priors.

The package \tpack{mglmn} is not Bayesian and it uses the Akaike Information Criterion (AIC).

\section{Selected Packages}
The {\tt R} packages \tpack{BayesFactor}, \tpack{BayesVarSel}, \tpack{BMS} and \tpack{mombf} provide functionalities to calculate and study the posterior distribution (\ref{postprob}) corresponding to some of the conventional priors described in Table~\ref{gTable}. The commands for such calculation are {\tt regressionBF()} in \tpack{BayesFactor}; {\tt Bvs()}, {\tt PBvs()} and {\tt GibbsBvs()} (for exhaustive enumeration, distributed enumeration and Gibbs sampling) in \tpack{BayesVarSel}; {\tt bms()} in \tpack{BMS} and finally {\tt modelSelection()} in the package \tpack{mombf}.

\paragraph{Prior inputs} The different conventional priors available in each package and the corresponding argument for its use are described in Table~\ref{tbl:pack-priors}.

\begin{table}[t!]
\begin{center}
{\small\scalebox{0.85}{
 \begin{tabular}{p{3.05cm}|lllll}

Package      						&\tpack{BayesFactor}	&\tpack{BayesVarSel}	&\tpack{BMS} 	&\tpack{mombf}\\\hline
Commands for model uncertainty          &{\tt regressionBF()} &{\tt Bvs(),PBvs(),GibbsBvs} & {\tt bms()} & {\tt modelSelection()}\\
\hline
\backslashbox[35mm]{Prior}{Argument}	&{\tt rscaleCont=}				  	&{\tt prior.betas=}	&{\tt g=} 	&{\tt priorCoef=}\\\hline
C1 									&-					&{\tt "gZellner"}	&{\tt "UIP"}&{\tt zellnerprior(tau=n)}\\
C2 									& -					&-					&{\tt "RIC"}&{\tt zellnerprior(tau=$p^ 2$)}\\
C3 									&-					&{\tt "FLS"}			&{\tt "BRIC"}&{\tt zellnerprior(tau=max(n,$p^2$))}\\
C4 								 	& -					&-					&{\tt "HQ"} 	&{\tt zellnerprior(tau=log(n))})\\
C5 									& -								&-		&{\tt "EBL"}	&-\\
\hline
R1 									& 1 		&{\tt "ZellnerSiow"}	&-		 	&-\\
R2 								 	& -					&-					&{\tt "hyper=a"}	&-\\
R3 (a=3)								&-					&{\tt "Liangetal"}	&-			&-\\ 	
R4 									&-					&{\tt "Robust"}		&- 			&-\\
\hline

\end{tabular}
}}
\end{center}
\caption{\small Priors for the parameters within each model. Main commands and corresponding modifying arguments for the different specifications for the hyper parameter $g$ (keys in column `Prior' refer to that in Table~\ref{gTable}) in conventional prior.}\label{tbl:pack-priors}
\end{table}

The implementation of the conventional priors in \tpack{mombf} have certain peculiarities that we now describe.
The priors for the common parameters, $(\alpha,\sigma)$, in \tpack{mombf} do not exactly coincide with (\ref{priorcommon}). In this package, the simplest model $M_0$ only contains the error term and hence $\alpha$ is not a common parameter. The more popular problem with fixed intercept examined in this paper (cf. (\ref{priorcommon})) is handled via the modifying argument {\tt center=TRUE} (given by default) which in turns is equivalent to a prior for $\alpha$ degenerate at its maximum likelihood estimate. This will, especially if $n$ is large enough, often lead to very similar results as with a flat prior on $\alpha$ but small differences could occur because in \tpack{mombf} the variability in this parameter is not taken into account. Also, for $\sigma^2$ this package uses an inverse gamma which has the non informative $\sigma^{-2}$ as a limiting density. Thus, differences among the two are expected to be negligible if the parameters in the inverse gamma are small (values of 0.01 are given by default). Another logical argument in {\tt modelSelection()} is {\tt scale}. If it is set to {\tt TRUE} the $y$'s and the $x$'s are scaled to have unitary variance. %Conventional priors are invariant to such transformations, but the prior used in \tpack{mombf} is not. Nevertheless, this issue should not have much impact on the results.
In this article we are fixing it to {\tt scale=FALSE} so that the data that enter in all the main functions exactly coincide.

\begin{table}[t!]
\begin{center}
{\small\scalebox{0.85}{
 \begin{tabular}{l|lllll}

Package      			 	&\tpack{BayesFactor}	&\tpack{BayesVarSel} 	&\tpack{BMS}		&\tpack{mombf}\\\hline
\backslashbox[35mm]{Prior}{Argument} 	
						 	&{\tt newPriorOdds(BFobject)=}			&{\tt prior.models=}	&{\tt mprior=}	&{\tt priorDelta=}\\\hline
$\theta=1/2$  &{\tt rep(1,2\^\ p)}					&{\tt "constant"}	&{\tt "fixed"} or {\tt "uniform" }&{\tt modelunifprior()}\\
%							&	($\pi=1/2$)					&($\pi=1/2$)			&$\pi$ (mirar) or $\pi=1/2$	&\\	\hline	
$\theta\sim$ Unif(0,1)			&-					&{\tt "ScottBerger"}	&{\tt "random"}	&{\tt modelbbprior(1,1)}\\
%							&					&($a=1$, $b=1$)		&ley and steel	&\\\hline
%User option 					&a vector with the prior  					&{\tt "User"}		&{\tt "customk"} and {\tt "pip"}&\\

%							&odds $P(M_\gamma)/P(M_0)$
%								&(\tpack{prior.probs})&	{\tt "mprior.size"}&\\

\hline
\end{tabular}
}}
\end{center}
\caption{\small {Most popular default priors} over the model space (see (\ref{eq:MSprior})) within the selected packages. For more flexible options see the text.}\label{tbl:pack-priorsMS}
\end{table}

All four packages are very rich and flexible regarding the choice of the prior over the model space, $Pr(M_\gamma)$. The access to the standard approaches is described in Table~\ref{tbl:pack-priorsMS}. Apart from these standard priors \tpack{BMS}, following the proposals in \cite{LeySteel09}, also allows for the use of a beta distribution for $\theta$ in (\ref{eq:MSprior}) by using {\tt mprior="random"} and modifying the argument {\tt mprior.size} to specify the desired expectation for the model prior distribution (the default option is $p/2$ hence providing the uniform prior on model size). Similarly the \tpack{mombf} package provides a beta prior for $\theta$ with parameter $(a,b)$ by setting the corresponding argument to {\tt modelbbprior(a,b)}. In \tpack{BayesVarSel} particular specifications of prior probabilities are available with {\tt mprior="User"} and a $p+1$ dimensional vector defined in {\tt priorprobs} which describes the prior probability, $Pr(M_\gamma)$, of a single model of each possible size (models of the same size are assumed to have the same prior probability). %over model size {divided by the number of models of each size}. %\rojo{**is this right?**} \azul{Comment for us:(see in blue the added text){\tt priorprobs} is not really the prior over model size but it should contain probabilities of single models of that size} %containing the prior probability, $Pr(M_\gamma)$, of a single model of each dimension (models of same dimensions are assumed to have the same prior probability).}

For illustration purposes consider the FLS dataset in \cite{LeySteel09} with $p=41$ potential regressors. These authors study the prior (\ref{eq:MSprior}) with $\theta\sim Beta(1,b=(41-\omega^\star)/\omega^\star)$ and $\omega^\star=7$, reflecting that, a priori, the expected number of regressors is $\omega^\star=7$. Such a prior can be implemented in \tpack{BMS} with {\tt mprior="random", mprior.size=7} and in \tpack{mombf}
with {\tt modelbbprior(1,34/7)}. In \tpack{BayesVarSel} the syntax is quite different and we have to
specify {\tt prior.models="User"} and
$$
{\tt priorprobs=dbetabinom.ab(x=0:41, size=41, shape1=1, shape2=34/7)/choose(41, 0:41)}.
$$

\paragraph{Summaries {and model averaging}} The result of executing the main commands for model uncertainty (see Table~\ref{tbl:pack-priors}) is an object describing, with a specific structure depending on the package, the posterior distribution (\ref{postprob}). For ease of exposition suppose the object created is called {\tt ob}. We compare here the different possibilities to summarize this distribution under each package. This is illustrated in the Supplementary Material which shows the different ways of summarizing the results for each package using one of the studied data sets.

\begin{itemize}
	\item In \tpack{BayesFactor}, a list of the most probable models and their corresponding Bayes factors (to the null model) can be obtained with the command {\tt head(ob)} or {\tt plot(ob)} over the resulting object.

	\item In \tpack{mombf}, this list can be obtained with {\tt postprob(ob)} but now best models are displayed with their posterior probabilities. Additionally, inclusion probabilities (\ref{eq:inclprob}) are contained in {\tt ob\$margpp}. In the context of large model spaces, having a list with all the models sampled can be very useful so that the user may program his/her own needs, such as  model averaged predictions. Such a list is contained in binary matrix form in \tpack{mombf} in {\tt ob\$postSample}. {To obtain model averaged estimates we also have the command {\tt rnlp} which produces posterior samples of regression coefficients (from which it is easy to obtain any $\hat{\Lambda}$ in (\ref{eq:bma}) that relates to coefficients).} %Advanced users can program routines to obtain model averaged predictions as the list of all sampled models is accessible.}

	\item In \tpack{BayesVarSel} most probable models and their probabilities are viewed printing the object created, {\tt ob}, while {\tt summary(ob)} displays a table with the inclusion probabilities, the HPM and the MPM (see Subsection \ref{Sec:Summaries}). The posterior distribution of the model size (\ref{eq:ppdim}) is in {\tt ob\$postprobdim} which can be graphed with {\tt plotBvs(ob,option="d")}. Plots of several measures of the joint importance of two covariates (e.g. joint inclusion probabilities) can be visualized as an image plot with {\tt plotBvs(ob, option="j")}. All models visited are saved in the matrix {\tt ob\$modelslogBF} which, in the last column, have the Bayes factors of each model in log scale.

	\item In \tpack{BMS} the top best models with their probabilities are displayed using {\tt topmodels(ob)}, that can also be plotted with {\tt image(ob)}. A {\tt summary(ob)} of the resulting object also prints the posterior of the model size (\ref{eq:ppdim}) that can be plotted with the command {\tt plotModelSize(ob)}. {Printing {\tt ob} displays a table with model averaged estimates of regression coefficients, namely their expected posterior mean and standard deviation (column Post Mean and Post SD respectively).} Interestingly, it is possible to compute predictions with the commands {\tt predict(ob)} (expected predictive mean a posteriori) and {\tt pred.density(ob)} (mixture predictive density based on a selected number of best models). This package does not save all the models visited but only a (necessarily small) list of the best models sampled in {\tt ob\$topmod} expressed in hexadecimal code.
	
	\end{itemize}

\paragraph{Numerical methods}
Exhaustive enumeration can be performed with \tpack{BayesFactor}, \tpack{BayesVarSel} (command {\tt Bvs()}) and in \tpack{BMS} (modifying argument {\tt mcmc="enumerate"}).

When $p$ is larger, exhaustive enumeration is not feasible and this is solved in \tpack{mombf}, \tpack{BayesVarSel} and \tpack{BMS} by providing specific routines to approximate the posterior distribution in such big model spaces. In summary, all three packages implement the strategy i) briefly described in Section~\ref{secNM} with the following peculiarities.
The packages \tpack{mombf} and \tpack{BayesVarSel} implement the same Gibbs sampling scheme. A minor difference between both is that frequency-based estimates of inclusion probabilities in \tpack{mombf} are refined using Rao-Blackwellization. The methods programmed in \tpack{BMS} are also MCMC strategies to explore the posterior distribution which can be of the type birth and death (modifying argument {\tt mcmc="bd"}) or a reversible jump ({\tt mcmc="rev.jump"}). There is an important difference between the algorithms in \tpack{mombf}, \tpack{BayesVarSel} and in \tpack{BMS}. While in each MCMC step the inclusion/exclusion of {\em all} $p$ covariates is sampled in \tpack{mombf} and \tpack{BayesVarSel} only one is sampled in \tpack{BMS}.

% \rojo{**the latter is a bit unclear to me**}\azul{removed}

\section{Performance in selected datasets}
To compare the selected packages two different scenarios have been considered:
\begin{itemize}
\item Exact scenario: data sets with small $p$ and hence all the models can be enumerated.
\item Sampling scenario: data sets with moderate to large $p$ where only a small proportion of models can be explored.
\end{itemize}
%For each scenario we consider two different real data sets.

As we previously mentioned, \tpack{mombf} cannot be considered in the exact scenario nor can \tpack{BayesFactor} be considered in the sampling scenario. Ideally, we should compare all possible packages (in each setup) under the same prior. Thus, Table~\ref{tbl:pack-priors} indicates which comparisons are possible. We compared \tpack{BayesFactor} with \tpack{BayesVarSel} using the Zellner-Siow prior (labelled as R1) while we compared \tpack{mombf} and \tpack{BMS} and  \tpack{BayesVarSel} using the UIP (C1). In all cases, the constant prior over the model space was used.

As expected, all four packages produced very similar results in the analyzed datasets. Hence, the question of comparing them reduces basically to comparing computational times and the availability, clarity and organisation of the output.

For the computational comparisons to be fair all the calculations have been done on an iMac computer with Intel Core i5, 2.7 GHz processor. The code used to compute results provided here is publicly available at  www.uv.es/fordela.

%\rojo{**how about comparing the availability, clarity and organisation of the output?**} \azul{Comment for us: we have attempted to provide a response to this but beyond what is said in the paragraph `summaries' of the previous section seems to be a little bit boring and repetitive. We then opted for including an appendix where something like a session with the different packages is reproduced. But we are not either convinced with this solution. If you have any specific suggestion in this aspect we would be delighted to work on it.}

\subsection{Exact Scenario}
We considered two data sets that we briefly describe.

\paragraph{US Crime Data.} The US Crime data set  was first studied by \cite{E73} and is available from R-package {\tt MASS} \citep{MASS}. This data set has a total of $n=47$  observations (corresponding to states in the US) of $p=15$ potential covariates aimed to explain the rate of crimes in a particular category per head of population.
\paragraph{Returns to schooling} This data set, used by \cite{ToLi04} and \cite{LeySteel12},  concerns returns to education. As these are microeconomic data, the number of potential observations is much larger. In particular we have a response variable: the log of hourly wages recorded for $n = 1190$ white males in the US in 1990, and a total of  $p=26$ possible regressors.
\mbox{}\\

For both scenarios we directly compare the time needed to exactly calculate the posterior distribution with \tpack{BayesVarSel} and \tpack{BMS} using the C1 prior for the parameters and the uniform prior (with fixed $\theta=1/2$) for the model space. These times are presented in Table~\ref{tbl:resultsExact}. The results clearly indicate that \tpack{BayesVarSel} is more affected by the sample size since it performs better than \tpack{BMS} for the Crime data set ($n=47$) but not for the returns to schooling application ($n=1190$). {The Bayes factors depend on the data only through the sum of squared errors and we know that \tpack{BayesVarSel} computes this statistic from scratch for each model and, thus, the $n$ matters in that calculation. Hence a likely reason for the differences in computational time between the two packages when $n$ increases would be that the algorithm in \tpack{BMS} exploits reduction by sufficiency and optimally updates when a variable is added/dropped from the current model.} %\rojo{**can we intuitively explain why this is the case?***}

The comparison between \tpack{BayesFactor} and \tpack{BayesVarSel}, now using the R1 prior, is summarized in the same table for the Crime data set where we can clearly see that \tpack{BayesFactor} is outperformed by \tpack{BayesVarSel}.

% latex table generated in R 3.1.3 by xtable 1.7-4 package
% Sat Dec 12 17:14:58 2015
\begin{table}[h!]
\centering
\begin{tabular}{lcccc}
  Data set & Prior & \tpack{BMS}  & \tpack{BayesVarSel} & \tpack{BayesFactor}\\ \hline
Crime  $p=15$& C1 unif & 3.22 secs & 0.35 secs & -\\
Returns to schooling  $p=26$& C1 unif & 1.83 hrs &  11.24 hrs & - \\
Crime  $p=15$& R1 unif & - & 1.4 secs & 12.73 mins\\
%Returns to schooling  $p=26$& R1 unif & - &  XXX hrs & ??? \\
  \hline
\end{tabular}
\caption{\small Computational times in exact scenario.}\label{tbl:resultsExact}
\end{table}

Table 4 also illustrates the large difference in computational cost between an exhaustive analysis with $p=15$ covariates (where $\cal M$ has $2^{15}=32,768$ models) and $p=26$, leading a model space with 67 million models, which is about 2000 times larger. Computational cost goes up by a factor of about 2000 for \tpack{BMS}, which is therefore roughly linear in the size of model space, and thus seems virtually unaffected by the number of observations $n$. {This is a consequence of how the statistics are computed within each package, as commented above.}%\rojo{**can we explain this, perhaps linked to the behaviour of \tpack{BayesVarSel}?**}\azul{Sure: is the same explanation. done}

\subsection{Sampling Scenario}
We considered here three data sets.

\paragraph{Ozone.} These data were used by \cite{CasMor06}, \cite{BerMol05} and \cite{Ga-DoMa-Be13} and contain $n = 178$ measures of ozone concentration in the atmosphere with a total of $p=35$ covariates. Details on the data can be found in \cite{CasMor06}.

\paragraph{GDP growth.} This dataset is larger than Ozone with a total of $p=67$ potential drivers for the annual GDP growth per capita between 1960 and 1996 for $n = 88$ countries. This data set is also used in \cite{SDM} and revisited by \cite{LeySteel07}.

\paragraph{Boston housing.} This dataset was used recently in \cite{SchCho13} and contains $n=506$ observations of $p=103$ covariates formed by the 13 columns of the original data set, all first order interactions and a squared version of each covariate (except for the binary variable {\tt CHAS}).\\

\begin{figure}[t!]
\centering
\includegraphics[width=0.9\textwidth]{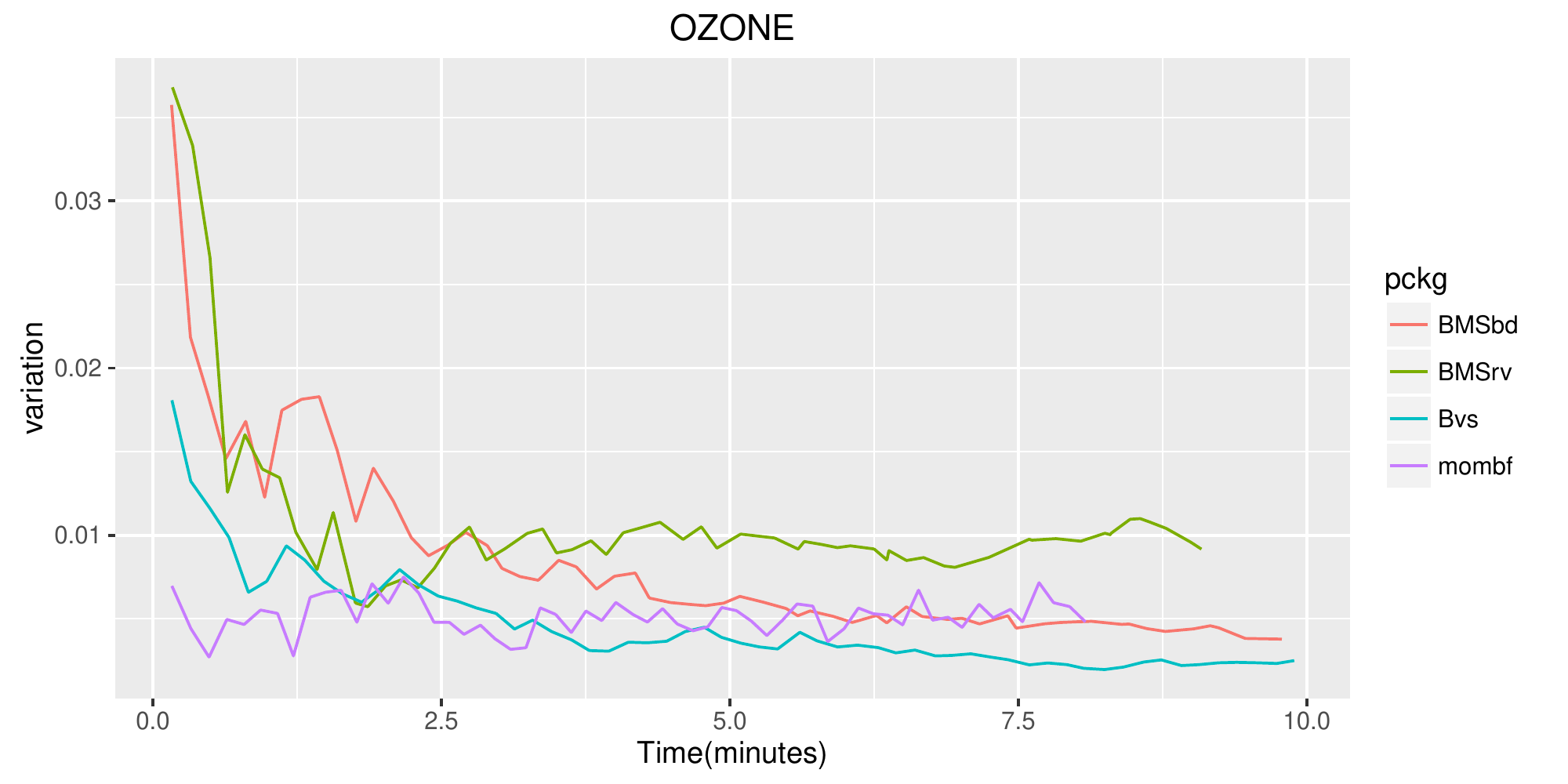}
\caption{\small Ozone dataset: maximum difference with the real inclusion probabilities ($\Delta_t$) as a function of computational time. BMSbd (BMSrv) stands for the birth/death (reversible jump) algorithm in \tpack{BMS}; Bvs for \tpack{BayesVarSel} and momfb for \tpack{mombf}.}\label{fig:exact}
\end{figure}

For the Ozone dataset, exact inclusion probabilities, (\ref{eq:inclprob}), are reported in \cite{Ga-DoMa-Be13} for the C1 prior. These are the result of an intensive computational experiment aimed at comparing different searching methods. These numbers allow us to define a simple measure to compare the computational efficiency of the different packages. For a given computational time, $t$, we can compute
$$
\Delta_t=\max_{i=1,\ldots,p}\, |\widehat{Pr}_t(\gamma_i=1\mid\n y)-Pr(\gamma_i=1\mid\n y)|,
$$
where $\widehat{Pr}_t(\gamma_i=1\mid\n y)$ is the estimate of the corresponding PIP at time $t$ provided by the package. Clearly, the faster $\Delta_t$ approaches zero,  the more efficient is the package. In Figure~\ref{fig:exact} we have plotted $\Delta_t$ for \tpack{mombf} and \tpack{BayesVarSel} respectively and the two algorithms in \tpack{BMS}.

All four approaches behave quite satisfactorily, providing very reliable estimates with a small computational time (a maximum discrepancy with the exact values of 0.01 in less than 2.5 minutes). It seems that \tpack{BayesVarSel} is slightly more efficient than the rest while the reversible jump implemented in \tpack{BMS} is less efficient. The apparent constant bias in \tpack{mombf} is possibly due to the difference in the prior actually implemented that we have already described.

\begin{figure}[t!]
\centering
\includegraphics[width=0.9\textwidth]{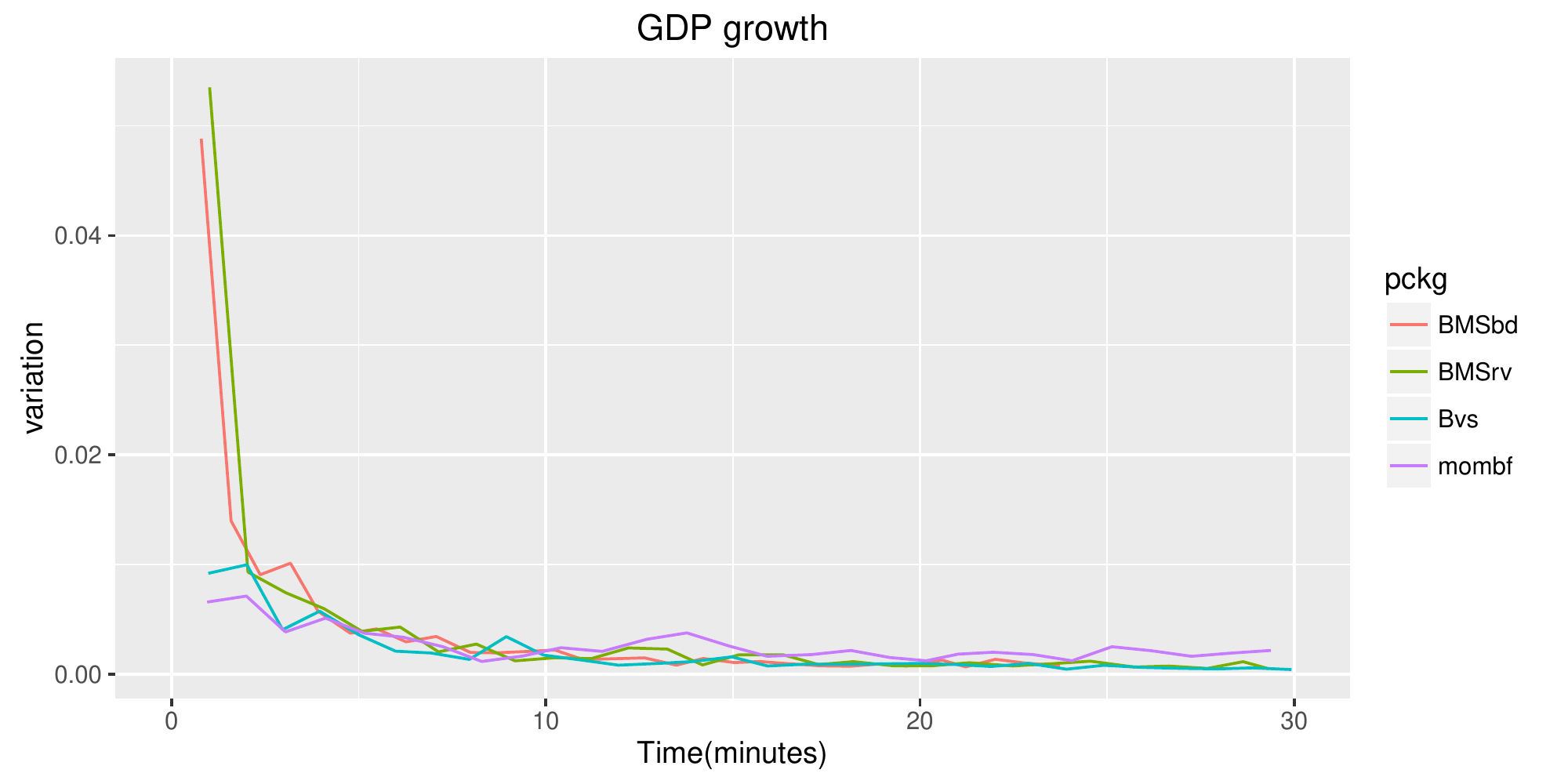}\\
\includegraphics[width=0.9\textwidth]{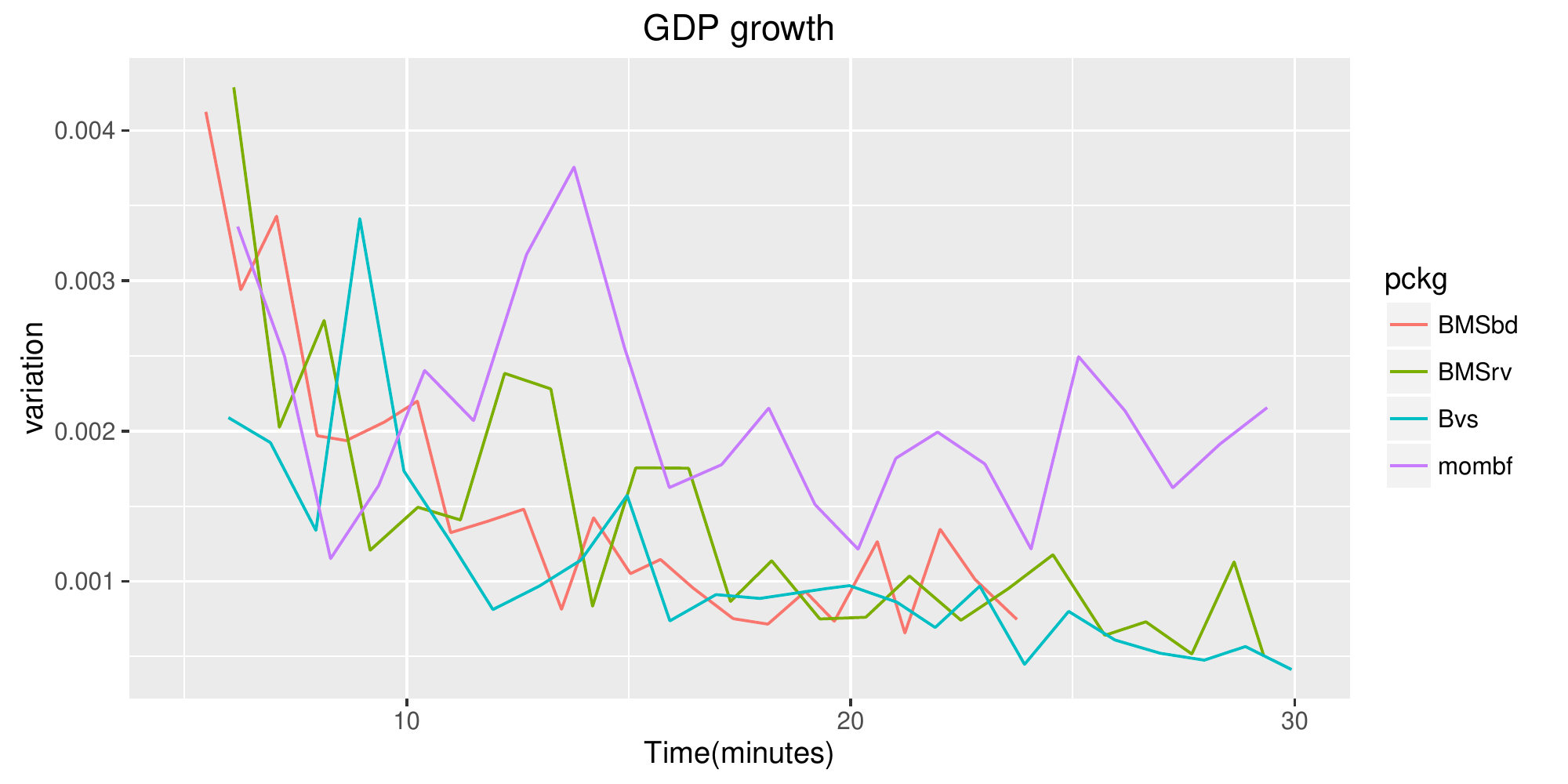}
\caption{\small GDP growth data: variations in PIP ($\Delta_{t,t-dt}$) as a function of computational time with $dt=60$ seconds (starting after the burning period). Both figures represents the same functions and just differ in that the figure below describes the behaviour after 5 minutes of computation.} %Labels defined in the caption of Figure~\ref{fig:exact}.}
\label{fig:variationGDP}
\end{figure}

\begin{figure}[t!]
\centering
\includegraphics[width=0.9\textwidth]{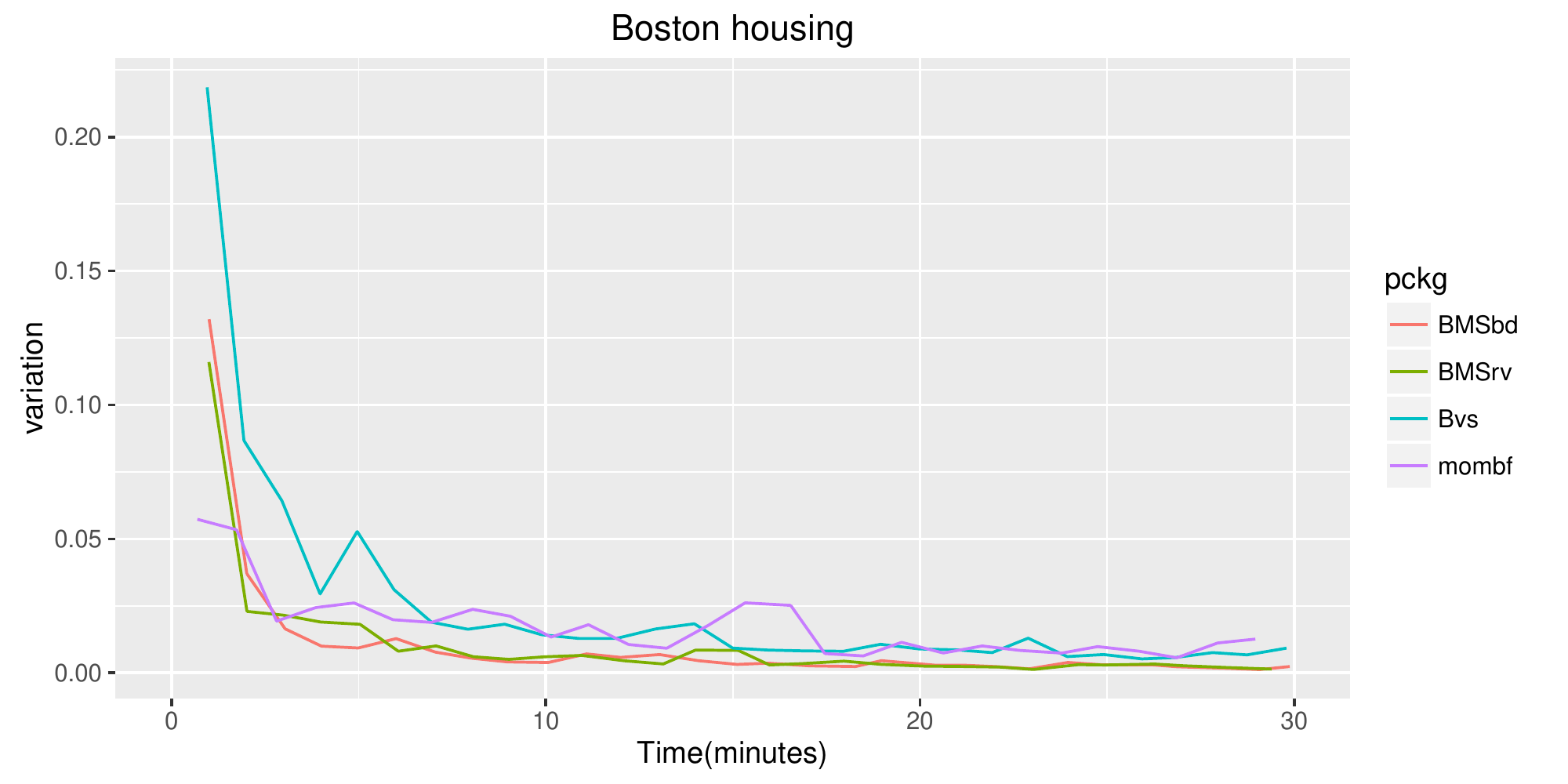}\\
\includegraphics[width=0.9\textwidth]{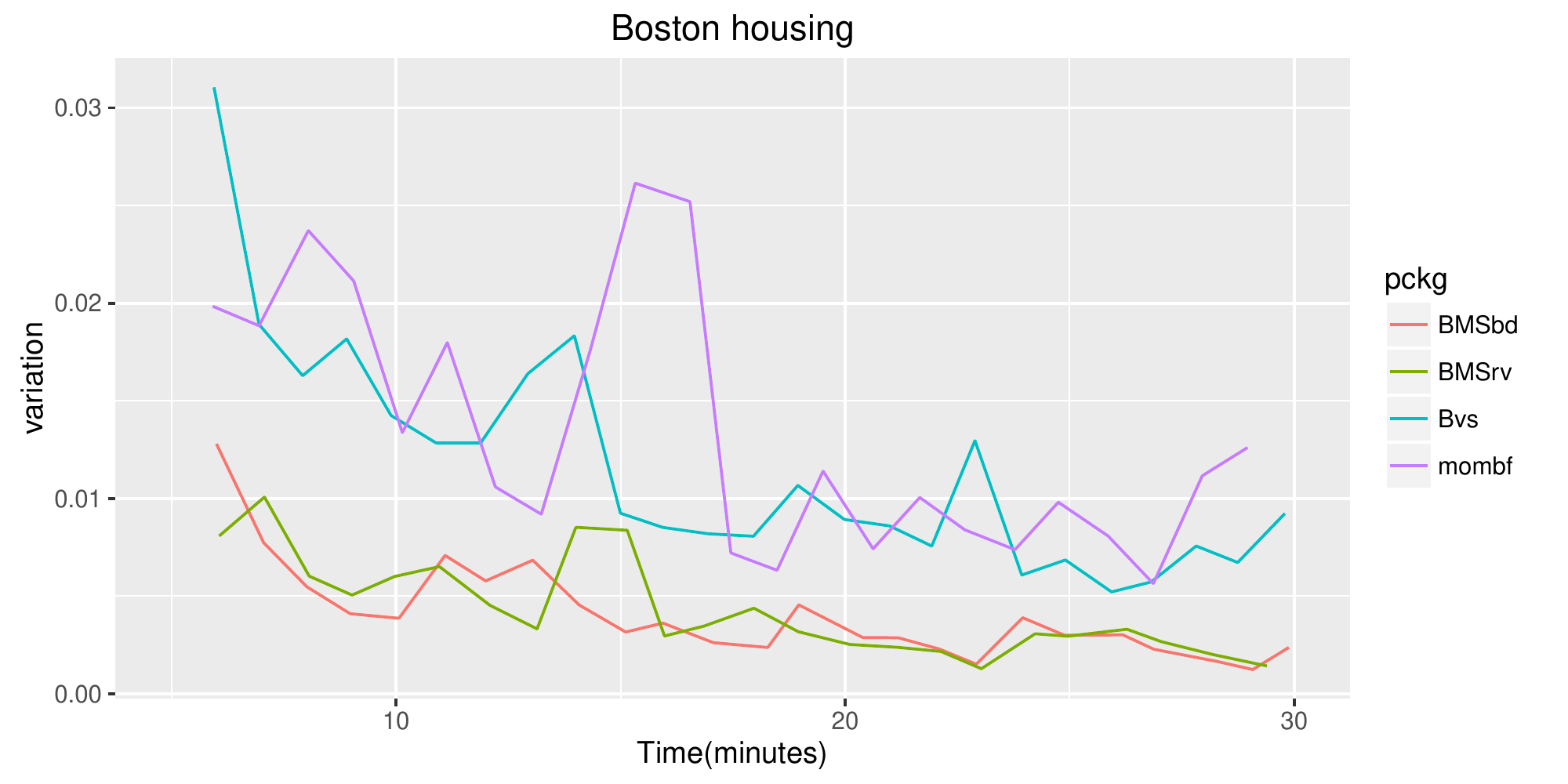}
\caption{\small Boston Housing data: variations in PIP ($\Delta_{t,t-dt}$) as a function of computational time with $dt=60$ seconds (starting after the burning period). Both figures represents the same functions and just differ in the axis represented (the figure below details the behaviour of the routines after 5 minutes of computation).}% Labels defined in the caption of Figure~\ref{fig:exact}.}
\label{fig:variationBH}
\end{figure}

In the GDP growth and the Boston Housing examples, we cannot compute $\Delta_t$ simply because the PIP's are unknown. Nevertheless, we observe that for a sufficiently large computational time, all packages converged to almost identical PIP's. Hence, and even in the unlikely case that none of them were capturing the `truth' it seems that the fairest way to compare the packages is computing time until `convergence'. This is what we have represented in Figures~\ref{fig:variationGDP} and \ref{fig:variationBH} where the $y$-axes display the difference between estimates at consecutive computational times, {\it i.e.}
$$
\Delta_{t,t-dt}=\max_{i=1,\ldots,p}\, |\widehat{Pr}_t(\gamma_i=1\mid\n y)-\widehat{Pr}_{t-dt}(\gamma_i=1\mid\n y)|,
$$
where $dt=60$ seconds was used and we have verified that PIPs converge. %\rojo{**maybe mention $dt$ used; if we take $dt$ small enough and all differences would be of the same sign, there could still be substantial "drift" in the results... I am assuming signs are alternating***}\azul{Done. Comment: Anabel has launched again the routines. We observe that pips are reasonably converging.}

In the GDP growth data set, we can not find big differences in the performance of all four approaches and all of them behave, again, very satisfactorily. It seems that the procedure implemented by \tpack{BayesVarSel}  tends to 0 faster than the rest of algorithms while the performance of \tpack{mombf} manifests more variability, likely due to the Rao-Blackwellization way of computing the results.

In the Boston Housing problem the package \tpack{BayesVarSel} is clearly penalized (with respect to the GDP growth data) by the large number of observations hence having a slower convergence.

For both examples, the inclusion probabilities obtained with each package after 30 minutes of computations (after the burning period)  differ, at most, in the second decimal number.
Finally, both plots are not affected by the difference in the prior implemented (as each method compares with itself). % and still \tpack{mombf} convergences slightly later than the others.

\section{Other Features}
Besides the characteristics analysed so far (prior inputs, numerical methods and summaries), there are several other features of the packages that are potentially relevant  for the applied user. We list some here under three categories: the interface, extra functionalities and documentation.

\paragraph{The interface} In general all four packages have simple interfaces with quite intuitive syntaxes. One minor difference is that in \tpack{BayesVarSel} and \tpack{BayesFactor} the dependent and explanatory variables are defined with the use of {\tt formula} (hence inspired by well-known {\tt R} commands like {\tt lm}) while in \tpack{mombf} these are defined through the arguments {\tt y} and {\tt x}. In \tpack{BMS} the dependent variable should be in the first column of the data provided and the rest play the role of explanatory variables.

\paragraph{Extra functionalities}
\begin{itemize}
  \item {\em Fixed covariates}. By default only the intercept is included in all the competing models (cf. (\ref{theproblem})) in all packages (but recall this is handled in \tpack{mombf} via centering). There could be situations where we wish to assume that certain covariates affect the response and these should be always included in the analysis \citep[see, for instance][]{Camarero15}. Both \tpack{BMS} and \tpack{BayesVarSel} include this possibility in their main commands.

	\item {\em Main terms and interactions}. On occasion, it is convenient to conserve the hierarchy between the explanatory variables in the way that  interactions (or higher polynomial terms) are only included if the main terms are included \citep{Pei87}. In \cite{ChipHaWu97} this is called the ``heredity principle''. This would translate into a reduction of the model space. The package \tpack{BMS} accommodates this possibility through a modification of the sampling algorithm.
	
	\item {\em Model comparison}. A complementary tool to the BMA exercise would be comparing separately some of the competing models (e.g. comparing the HPM and the MPM). These type of comparisons can be performed in \tpack{BMS}, \tpack{BayesVarSel} and \tpack{BayesFactor}.
	
	\item {\em Convergence}. \tpack{BMS} includes several interesting tools to analyse the convergence of the sampling methods implemented.
	
	\item {\em Parallel computation}. \tpack{BMS}, \tpack{BayesVarSel} and \tpack{mombf} have facilities to perform computations in parallel.
	
%		\item {\em Predictions}. Several solutions to produce posterior predictions are available in \tpack{BMS} with the commands {\tt predict()} and {\tt pred.density()}. Advanced users can easily program all sort of routines for predictions with \tpack{mombf} and \tpack{BayesVarSel} as the list of all sampled models is provided. Predictions can be a very useful tool to run predictive checks (often using score functions) e.g. to compare various prior specifications.
	
  \end{itemize}

\paragraph{Documentation} The four packages come with a detailed help with useful examples. Further, \tpack{mombf} and \tpack{BMS} have a comprehensive {\em vignette} with additional illustrations and written more pedagogically than the help documentations.

The packages \tpack{BMS} and \tpack{BayesFactor} are documented in the websites associated with \cite{webBMS16} (\url{http://bms.zeugner.eu}) and \cite{webBF16} (\url{http://bayesfactor.blogspot.com.es}), respectively. %\rojo{**mention URLs?**}\azul{done}.
These sites contain manuals as well as valuable additional information, especially to users less familiars with model uncertainty techniques.

\section{Conclusions and recommendations}

In this paper, we have examined the behaviour and the possibilities of various {\tt R}-packages available in CRAN for the purpose of Bayesian variable selection in linear regression. In particular, we compare the packages \tpack{BMS}, \tpack{BayesVarSel}, \tpack{mombf} and \tpack{BayesFactor}. It is clear that all packages concerned lead to very similar results, which is reassuring for the user. However, they do differ in the prior choices they allow, the way they present the output and the numerical strategies used. The latter affects CPU times, and, for example means that \tpack{BayesVarSel} is a good choice for small or moderate values of $n$, but  \tpack{BMS} is preferable when $n$ is large. The package \tpack{BayesFactor} can not deal with larger values of $p$ and seems relatively slow, thus is not  recommended  for general use. \tpack{mombf} uses a slightly different prior from the one we focus on here (and which is the most commonly used), but is relatively competitive and closely approximates the PIPs after a short run time, albeit with slightly more variability than \tpack{BMS} or \tpack{BayesVarSel}.

In practice, users may be interested in specific features, such as always including certain covariates, that will dictate the choice of package. On the basis of its performance, the flexibility of prior choices and the extra  features allowed, we would recommend the use of \tpack{BayesVarSel}  for small or moderate values of $n$, and of  \tpack{BMS} when $n$ is large.

\section*{Acknowledgments}
The authors would like to thank David Rossell for valuable comments on a preliminary version of this paper.
\bibliographystyle{chicago}
%\bibliography{BMAinR_complete}
\bibliography{../../bibliography/BMAinR_complete}

\end{document}